\documentclass[twocolumn,trackchanges]{aastex62}
\usepackage{amsmath}
\usepackage{multirow}
\usepackage{subfigure}

\newcommand{\pair}{\ensuremath{e^\pm}}

\newcommand{\msun}{\ensuremath{\rm{M}_\odot}}

\newcommand{\nickel}{\ensuremath{^{56}\mathrm{Ni}}}
\newcommand{\ergs}{\ensuremath{\rm{ergs}}}

\newcommand{\mlt}{\ensuremath{\alpha_{MLT}}}
\newcommand{\fover}{\ensuremath{f_{ov}}}
\newcommand{\wein}{\ensuremath{\sin^2\theta_{W}}}
\newcommand{\neu}{\ensuremath{\nu_{\rm{rate}}}}

\newcommand{\maxbh}{\ensuremath{\rm{M}_{\rm{BH,max}}}}
\newcommand{\mheint}{\ensuremath{\rm{M_{He,int}}}}

\newcommand{\mhe}{\ensuremath{\rm{M_{He}}}}
\newcommand{\mco}{\ensuremath{\rm{M_{\rm CO}}}}

\newcommand{\crate}{\ensuremath{^{12}\rm{C\left(\alpha,\gamma\right)^{16}O}}}

\newcommand{\code}[1]{\texttt{#1}}

\newcommand{\MESA}{\code{MESA}}
\newcommand{\STARLIB}{\code{STARLIB}}
\newcommand{\REACLIB}{\code{REACLIB}}
\newcommand{\NACRE}{\code{NACRE}}

\newcommand{\kms}{{\mathrm{km\ s^{-1}}}}

\DeclareRobustCommand{\Eqref}[1]{Equation ~\ref{#1}}

\DeclareRobustCommand{\eqref}[1]{\Eqref{#1}}

\begin{document}

\title{Mind the gap: The location of the lower edge of the pair instability supernovae black hole mass gap}

    
\author[0000-0003-3441-7624]{R. Farmer}
\affiliation{Anton Pannekoek Institute for Astronomy and GRAPPA, University of Amsterdam, NL-1090 GE Amsterdam, The Netherlands}

\author[0000-0002-6718-9472]{M.~Renzo}
\affiliation{Anton Pannekoek Institute for Astronomy and GRAPPA, University of Amsterdam, NL-1090 GE Amsterdam, The Netherlands}

\author[0000-0001-9336-2825]{S.~E.~de~Mink}
\affiliation{Center for Astrophysics | Harvard \& Smithsonian, 60 Garden Street, Cambridge, MA 02138, USA}
\affiliation{Anton Pannekoek Institute for Astronomy and GRAPPA, University of Amsterdam, NL-1090 GE Amsterdam, The Netherlands}

\author[0000-0002-0338-8181]{P.~Marchant}
\affiliation{Center
for Interdisciplinary Exploration and Research in Astrophysics (CIERA) and Department of Physics and Astronomy, Northwestern University, 2145 Sheridan Road, Evanston, IL 60208, USA}

\author[0000-0001-7969-1569]{S.~Justham}
\affiliation{School of Astronomy \& Space Science, University of the Chinese Academy of Sciences, Beijing 100012, China}
\affiliation{National Astronomical Observatories, Chinese Academy of Sciences, Beijing 100012, China}
\affiliation{Anton Pannekoek Institute for Astronomy and GRAPPA, University of Amsterdam, NL-1090 GE Amsterdam, The Netherlands}

\correspondingauthor{R. Farmer}
\email{r.j.farmer@uva.nl}

\begin{abstract}
Gravitational-wave detections are now starting to probe the mass distribution of stellar-mass 
black holes (BHs). Robust predictions from stellar models are needed to interpret these. 
Theory predicts the existence of a gap in the BH mass distribution because of 
pair-instability supernova. The maximum BH mass below the gap is the result of 
pulsational mass loss. We evolve massive helium stars through their late 
hydrodynamical phases of evolution using the open-source \MESA{} stellar 
evolution code. We find that the location of the lower edge of the mass gap 
at 45\msun{} is remarkably robust against variations in the metallicity 
($\approx 3\msun$), the treatment of internal mixing ($\approx 1\msun$), stellar 
wind mass loss ($\approx 4\msun$), making it the most robust predictions for 
the final stages of massive star evolution. The reason is that the onset 
of the instability is dictated by the near-final core mass, which in turn 
sets the resulting BH mass.  However, varying \crate{} 
reaction rate within its $1\sigma$ uncertainties shifts the location of the 
gap between $40\msun$ and $56\msun$. We provide updated analytic fits for 
population synthesis simulations.  Our results imply that the detection of
merging BHs can provide constraints on nuclear astrophysics. Furthermore, 
the robustness against metallicity suggests that there is a universal 
maximum for the location of the lower edge of the gap, which is insensitive 
to the formation environment and redshift for first-generation BHs. This is 
promising for the possibility to use the location of the gap as a 
``standard siren'' across the Universe.
\end{abstract}


\section{Introduction}
\label{sec:introduction}

The detection of merging black hole (BH) binaries through gravitational 
waves \citep[e.g.,][]{abbott_2016_aa, ligo18b} has opened an observational window on the most
massive stellar BHs in the Universe. Stellar evolution theory predicts 
the existence of a gap in the BH mass distribution due to pair-instability evolution
\citep[][]{fowler:64, barkat:67, woosley:17}, and the current population 
of detected binary BHs are consistent with a 
lack of BHs with masses $\geq 45\msun$ 
\citep{abbott_2016_ab,abbott_2016_aa,abbott_2017_ac,abbott_2017_ah,abbott_2017_ai}. 
So far, the most massive BH found is the primary of GW170729, with a mass 
of $50.6^{+16.6}_{-10.2}\,\msun$ \citep{ligo18a}.
This object is at the edge of the theoretically predicted mass gap.
\cite{fishbach:17} showed that the existence of the gap and the maximum 
BH mass at its
lower edge can be significantly constrained with the detections expected 
during the third LIGO/Virgo observing run.
  
The existence of this pair-instability BH mass gap is expected because 
of the occurrence of pair-instability supernovae (PISN) which can completely disrupt
the progenitor star leaving no compact remnant behind 
\citep[][]{rakavy:67,fraley:68, woosley:02}.
However, it is the mass loss during the pulsations in a pulsational pair instability supernovae (PPISN) 
that set the lower edge of this PISN BH mass gap.
PPISN are predicted for stars slightly less massive than PISN progenitors, 
and they leave behind a BH, but only after having experienced several
episodes of pulsational mass loss, which reduce the mass of the final BH.

Here, we investigate how robust the location of the lower edge of the BH mass gap due to PPISN is
\citep{rakavy:67,fraley:68}, and in particular its lower boundary, i.e., 
how massive can the most massive BH below the gap be. Single stars with 
initial masses $100\,\msun\lesssim M_\mathrm{ZAMS}
\lesssim 140\,\msun$ (or equivalently final helium core masses of 
$32\,\msun\lesssim\mhe\lesssim60\,\msun$), are expected to
undergo pulsation pair instabilities (PPI) \citep[][Renzo et al subm.]{woosley:02,chen:14,yoshida:16, woosley:17}. This
instability results in a series of pulses, each removing mass
from the star.  Eventually, the core stabilizes, the pulses cease, 
and the star ends its evolution in
an iron core collapse (CC) most likely producing a
BH \citep{barkat:67, woosley:17}. 

More massive stars are fully disrupted instead of 
producing ever more massive BHs: for initial masses 
$140\,\msun\lesssim M_\mathrm{ZAMS}\lesssim260\,\msun$ (metallicity dependent), corresponding
roughly to final helium cores $60\,\msun\lesssim
\mhe \lesssim 140\,\msun$ \citep{heger02}, the first pulse is so violent that the
entire star is fully disrupted in a PISN \citep{woosley:02,heger:03},
without any BH remnant formed. For even higher initial masses, 
corresponding to final $M_\mathrm{He}\gtrsim130\msun$, the 
photodisintegration instability allows again for BH formation 
\citep[][]{heger:03}, closing the PISN BH mass gap from above.

From a population of binary BH mergers, we can determine their rate 
\citep{abbott_2016_ab}, and their
mass distribution \citep{cutler94,kovetz17}.
However, the time-scale for binary BHs to merge
can be of the order of giga-years \citep{paczynski67}. Therefore, even 
if determining the host galaxy is possible despite the limited spatial 
localization of binary BH mergers the local observed population of stars may have
formed later and hence have a different metallicity 
to that of the BH progenitor. This complicates 
estimating the rate of BH formation
\citep{Portegies00,dominik:12,abbott_2016_ab},
since this estimate requires knowing the star formation rate and metallicity evolution
of the Universe \citep{madau14,neijssel19,mapelli19}.

The maximum BH mass below the PISN gap however can be more easily determined \citep{ligo18b}
as it is independent of the rate of BH formation. We can thus use
it without knowing the metallicity dependent star formation rate of the Universe. 
In this study we explore how sensitive the maximum BH mass is to uncertainties in
the metallicity of the progenitors as well as other known uncertainties 
in our understanding of stellar physics.

In section \ref{sec:method} we describe the evolution of PPISN and PISN
while introducing our computational approach. We outline the parameter variations we consider in section \ref{sec:parameters}.
In section \ref{sec:met} we discuss the sensitivity of the maximum BH mass
to changes in the metallicity of the stellar progenitors. Section \ref{sec:physics}
explores how uncertain the maximum BH mass below the PISN mass gap is because of uncertainties in
the assumed input physics. We discuss the implications of the maximum BH mass in 
section \ref{sec:implicat}. We conclude and summarize our results
in sections \ref{sec:desc} \& \ref{sec:conc}. 

\section{Evolution though the pulses}
\label{sec:method}

Using \MESA{}
version\footnote{This version is not an official release, but it is
publicly available from \url{http://mesa.sourceforge.net/}} 11123 \citep{paxton:11,paxton:13,paxton:15,paxton:18,paxton:19}, 
we evolve a series of single bare
helium cores until they undergo either PPI followed by a core 
collapse supernovae (PPISN) or the more violent pair instability that 
fully disrupts the star in a PISN. Input files necessary to 
reproduce this work and the resulting output files are made freely
available at \href{www.mesastar.org}{www.mesastar.org}\footnote{As well as \href{http://doi.org/10.5281/zenodo.3346593}{http://doi.org/10.5281/zenodo.3346593} }. 

Based on the results of 
\citet{marchant:18}, we evolve systems around the lower edge of
the PISN BH mass gap, 
with initial helium core masses between 30-105\,\msun. 
We chose to evolve bare helium cores as stars in this mass range are expected
to lose their hydrogen-rich envelope long before their death. 
This could happen either through binary interactions
\citep{kobulnicky:07,sana:12,almeida:17},
strong stellar winds \citep{vink:05, renzo:17},
LBV-like mass loss \citep{humphreys94},
opacity-driven pulsations in the envelope \citep{moriya15},
 or because of chemically
homogeneous evolution due to fast rotation 
\cite{maeder:00, yoon:06,demink:09, mandel:16b, marchant:16}. 

As stars evolve from the zero age helium branch (ZAHB)
they proceed by burning helium convectively in their core which encompasses $\sim90\%$ of the mass,
taking $\sim 10^{5}\,\rm{years}$. Once helium has been burnt in the core
convection ceases, leaving behind a carbon/oxygen (CO) core with an outer
helium burning shell surrounded by a helium-rich surface layer. 
For sufficiently massive cores
an inner region of the star will enter the pair instability region.
Due to dynamical instability from the production of \pair{} pairs softening the equation of state the
core begins contracting and heating up.
Eventually this region will heat up sufficiently to 
ignite the residual carbon and explosively ignite the oxygen \citep{fowler:64,rakavy:67,barkat:67}. 

This ignition will reverse the contraction and may generate an outwardly propagating 
pulse, if the star was sufficiently massive. As this pulse propagates 
outwards the inner region of the star expands and 
cools. Once the pulse reaches the surface, it steepens into a shock
wave which can then accelerate material
beyond the escape velocity. This removes between a few tenths and a few tens of solar masses of 
material in a pulsational mass loss episode (PPI) 
\citep{yoshida:16,marchant:18,woosley19}. 
Some stars will undergo ``weak'' pulsations, these stars 
undergo PPI instabilities but do not drive 
a shock sufficient to remove mass \citep{woosley:17, marchant:18}. 
To focus on the impact that this process has on the BH masses, in 
this study we define only systems which can drive mass
loss as undergoing a pulse. We define weak pulses as ones only able to drive 
small amounts of mass loss $\approx0.1\msun$ per pulse, while
strong pulses drive up to several tens of solar masses lost per pulse.
The star then contracts and cools either via neutrinos or in the most massive 
cores undergoing PPIs via radiative energy losses \citep{woosley:17,marchant:18}.
This cycle of contraction and ignition can occur multiple times. 

This contraction and expansion process is hydrodynamical in nature, generating 
multiple shocks. To model these shocks we uses \MESA's HLLC contact solver \citep{Toro1994,paxton:18} 
However for computational
reasons we do not use the HLLC solver while the star is in hydrostatic
equilibrium. Instead, only as the star evolves away from hydrostatic equilibrium
do we switch to using the HLLC solver.
We then follow the hydrodynamics through the ignition and 
expansion of the star. 
Once all secondary 
shocks have reached the surface, we excise any material that has a velocity greater
than the escape 
velocity
\citep{yoshida:16,marchant:18}. We then create a new stellar model with the 
same mass, chemical composition, and entropy as the previous model had 
(minus the excised material). At this point we 
switch back to using \MESA's hydrostatic solver 
as the star can be approximated as being in hydrostatic
equilibrium. This model is then evolved until the next pulse, where this process
repeats, or on to core collapse, which is defined when any part of the 
star infalls with $v>8\,000\,\kms$. Stars which undergo a PISN are evolved 
until all stellar material becomes unbound. 

We define the time just before a pulse to be when the pressure weighted  
integral of $\langle\Gamma_1\rangle<4/3$ \citep{stothers:99,marchant:18}.
A special case occurs once the core temperature ($\rm{T_c}$) exceeds $\rm{T_c} > 
10^{9.6}\rm{K}$, when we continue using the HLLC solver as the
star is approaching CC. During the 
hydrodynamical phases we turn off mass loss from winds. Given the short
amount of physical time spent by our models during the hydrodynamical
phase of evolution and the typical wind mass
loss rates of $\approx10^{-5}\,\msun\ \rm{yr}^{-1}$, this does not influence significantly 
the final BH masses.

We define the mass of the BH formed to be the 
mass enclosed with a 
binding energy $>10^{48}\ergs{}$ (e.g., \citealt{nadezhin:80,lovegrove:13,fernandez18})  
and velocities less than the escape velocity, measured at iron core collapse.
Stars which undergo a PISN are 
expected to be fully disrupted and thus leave no remnant behind. The 
final BH mass may depend on the mass of neutrinos lost during 
the collapse, assuming they are not accreted into the BH \citep{coughlin:18}.
 Without a 
fully consistent theory for BH formation, we use this simple value based on the 
binding energy, which 
provides an upper limit on the BH mass. This value of $10^{48}\ergs{}$
is a conservative estimate for the minimum energy released when a star collapses into a BH,
due to neutrino emission \citep{nadezhin:80,lovegrove:13,fernandez18}.
In general this limit is $\approx0.01\msun$ 
smaller 
than the total mass of bound material at core collapse. We 
define the location in mass of the CO core at the end of 
core helium burning where $X\left(^{12}\rm{C}\right) > 0.01$ and 
$X\left(^{4}\rm{He}\right) < 0.01$.

\section{Choice of parameters}
\label{sec:parameters}

There are many uncertain ingredients in the modelling of 
stars. These can either be algorithmic
parameters that are insufficiently
constrained by 
experiments or observations (e.g., convective mixing)
or physical quantities that can 
only be measured in regimes which are
much different than the stellar case
and require complicated and
uncertain extrapolation for their 
applications to stars (e.g., nuclear reaction rates). 
Thus we model a range of systems, with differing environmental, physical, and
numerical parameters to test the sensitivity of our results to these
parameters.

\subsection{Metallicity}

Since LIGO has the ability to detect stellar mass BH mergers out to red-shifts $\approx1$, for stellar
mass BHs, and the potential for the progenitor stars to come from even earlier epochs
it can thus probe the history of star formation across the Universe \citep{ligo18b}. 
Thus we evolve a series of models with varying metallicities ($\rm{Z}$) 
between $10^{-5}$ and $3\times10^{-3}$. Metallicity primarily effects the evolution
of a helium core by varying the 
amount of mass-lost via winds (see section \ref{sec:winds}), due to the wind-lost prescriptions strong 
dependence on the metallicity \citep{vink:01,Mokiem+2007a}.
The lower limit results in stars that do not not lose
any significant amount of mass though winds.
The upper limit is set by the requirement for us to be able to robustly model the
 PPISN and PISN region. 
The upper limit used is comparable to the physical upper limit found in \citet{langer:07} 
for H-rich PPISN progenitors.
At higher metallicities stars lose sufficient
mass that they do not enter the pair instability region and instead
evolve in hydrostatic equilibrium though carbon, oxygen, and silicon burning and then 
undergo direct collapse, likely forming a BH when they try to burn iron.
Our fiducial metallicity, when varying other physics parameters,
is $\rm{Z=10^{-3}}$.

\subsection{Wind mass loss}\label{sec:winds}

The total mass a star loses during its evolution plays a critical 
role in the fate of the star,
however just as important is how and when it loses the mass. Mass 
loss via winds is 
not self-consistently solved in 1D stellar evolution models, but instead, is set by a mass
loss prescription and that functional form can have a large impact
on the star's evolution \citep{renzo:17}.

We investigate three different wind mass loss algorithms, each having a different 
dependence on the stellar properties: the prescription of
\citet{hamann:82, hamann:95, hamann:98} (H); the prescription  of \citet{nugis:00} (N\&L);
the prescription of \citet{tramper:16} (T); as well as no mass loss ($\dot{\rm{M}}=0$). 

The helium cores we investigate
have surface luminosities $\approx10^6\,\rm{L_\odot}$, which is at the upper edge of 
currently known Wolf-Rayet stars used to derive these prescriptions. Thus we also 
append a free
scaling factor $\eta$ to test possible uncertainties in our knowledge of 
mass loss rates in high luminosity
helium cores. This free scaling parameter can be related to the inhomogeneities in
the wind structure (so-called ``clumpiness'') with
$\eta=\sqrt{\langle\rho^2\rangle/\langle \rho\rangle^2}$, where $\rho$
is the wind mass density, and the angle brackets indicate the spatial
average over the stellar surface. We vary $\eta$ between 0.1 and 1.0 \citep{smith:14}, 
with our fiducial wind 
being the (H) rate with $\eta=0.1$ \citep{yoon:10}. 
We assume a value of $\rm{Z}_{\odot}=0.014$ \citep{asplund:09}.

\subsection{Neutrino physics}

The evolution of massive stars is governed by neutrino losses, as the star
evolves to higher core temperatures
and densities the rate of thermal neutrino losses increases. 
Stars undergoing pulsational instabilities are also sensitive to the neutrino cooling rates, 
as due to the generation of \pair{} they produce copious amounts of neutrinos 
from their annihilation which leads to the core cooling. The stronger the cooling,
the more energy is required from nuclear burning to overcome these loses.

\MESA{} implements the analytic fits to neutrino losses from \citet{itoh96} 
for pair, photo,
plasma, bremsstrahlung and recombination neutrino processes. 
These fits have a quoted fitting errors of $\approx 10\%$ 
for pair, $\approx 1\%$ for photo, $\approx 5\%$ for plasma, 
$\approx 10\%$ for recombination, neutrinos compared to 
the detailed calculations for the regions where these processes 
are dominant \citep{itoh96}. Outside of the dominant 
regions the error increases rapidly. Bremsstrahlung 
neutrino losses have no quoted error, thus we assume a $\approx 10\%$ error, similar
to the other processes. 
We test the uncertainity due to this 
fitting error by
varying the neutrino rates by increasing (decreasing) the neutrino loss rate
by multiples of the quoted fitting error. While 
\citet{itoh96} states that the analytic fits will generally 
under predict the true value, we test both over and under estimates for completeness. 

A second important factor for the rate of neutrino loss in stars is the 
Weinberg angle, or the weak mixing angle from the 
Weinberg\textemdash Salam theory of the electroweak interaction \citep{weinberg67,salam68}. 
In the analytical fits of \citet{itoh96}, the Weinberg 
angle sets the relative rate of neutrino production between neutral current reactions
and charged current neutrino reactions. Increasing the Weinberg angle 
increases the neutrino cooling rate, by increasing the fraction of charged current
reactions.
While individual 
measurements of the Weinberg angle have small
quoted uncertainties, there is an systematic offset between different 
values which is larger 
than the quoted uncertainties.
Thus we model three values for the Weinberg angle 0.2319 \citep{itoh96}
(our fiducial value), 0.23867 \citep{erler05}, and 0.2223 \citep{codata14}. 
Over the range of Weinberg angles considered here, we find the neutrino rates vary by up to
$\approx3\%$, with the greatest change being in the pair-creation region.

\subsection{Mixing}

Convection inside a star is a difficult process to model 
\citep{bohmvitense:58,canuto96,meakin07}, especially during dynamical phases
of a star's evolution \citep{Chatzopoulos14,Chatzopoulos16}. 
Thus, we take a simpler approach and restrict ourselves to testing uncertainties 
within the framework of mixing length theory (MLT).
Specifically, we test the MLT's \mlt{} efficiency parameter
between 1.5 and 2.0, with 2.0 being our fiducial value. While this may not capture
the true uncertainty due to convection, it can provide bounds on the result. 
We use the prescription of convective velocities from \citet{marchant:18}
to limit the acceleration of convective regions.

At the convective boundaries we assume convective overshoot mixing with an 
exponential profile. This
is parameterized into two terms, $\fover$ and $f_0$, the first term dictates 
the scale height of
the convective overshoot, in units of the pressure scale height. 
The second term dictates the starting point inside the convective boundary
from where the overshoot begins, in pressure scale heights \citep{paxton:11}. We assume the 
value of $f_0=0.005$, and vary
$\fover$ between 0.0 (no overshooting) and 0.05, with $\fover=0.01$ being our fiducial
value. 

\subsection{Nuclear physics}

Nuclear reaction rates are highly sensitive to the 
temperature of at which the reaction occurs, and due to this
sensitivity the uncertainty in the rate is also highly 
temperature dependent \citep{iliadis_2010_ab,iliadis_2010_aa,longland_2010_aa}.  
Varying
nuclear reaction rate within its known uncertainties has 
been shown to a have large impact on the
stellar structure of a star \citep{hoffman99,iliadis02,fields16,fields18}.

We vary several nuclear reaction rates between their $\pm1\sigma$
uncertainties with data from \STARLIB{} \citep{sallaska13}. 
\MESA's default
rate set is a combination of \NACRE{} \citep{angulo99} and \REACLIB{} \citep{cyburt10}. 
To sample the rates, 
we take the median value from \STARLIB{} and by taking
the uncertainty on a rate to be a log normal distribution we can 
compute both an upper and lower rate
(given by $\pm1\sigma$) to cover 68\% of the rate's probability distribution. 
These bounds vary as a function of temperature
reflecting the varying uncertainty in the underlying 
experimental data. 
When sampling the rates, we vary only one rate at a time,
with the reminder of the rates being taken from \NACRE{} and \REACLIB{}.
Correlations between rates can impact the structure of a star 
and deserve further study \citep{fields16,fields18}.

We test variations in three rates; $3\alpha$ is the triple alpha reaction, 
$\rm{C12\alpha}$ is the \crate{}
reaction, and $\rm{O16\alpha}$ is the 
$\rm{^{16}O\left(\alpha,\gamma\right)^{20}Ne}$ reaction.
We choose to vary only a few rates over their $1\sigma$ uncertainties to 
limit the computational cost.

We also investigate the effect of changing the nuclear network used, 
which can have a large impact on the evolution of massive stars, 
due to changes in both which isotopes and which reactions are followed \citep{farmer16}.
By default we use the \code{approx21.net} which follows alpha 
chain reactions from carbon and iron, and includes compound reactions to follow
$\left(\alpha,\rm{p}\right)\left(\rm{p},\gamma\right)$ reactions (which assumes
that the intermediate isotope is in a steady state equilibrium) \citep{Timmes99,Timmes00}. 
We also evolve models with both
\code{mesa\_75.net}, which has 75 isotopes up to $^{60}\rm{Zn}$, 
and \code{mesa\_128.net}, which has 128 isotopes up to $^{60}\rm{Zn}$, including
more neutron rich nuclei than the \code{mesa\_75.net} network, which do not include
any compound reactions. 

\subsection{Other physics}

\MESA{} is built upon a range of other physics, which we do not vary here 
but which can provide other uncertainties in the modelling of massive stars.
\MESA's equation of state (EOS) for massive stars is a blend of the OPAL \citep{Rogers2002} and HELM
\citep{Timmes2000} EOSes. 
Radiative opacities are primarily from OPAL \citep{Iglesias1993,Iglesias1996}, 
with low-temperature data from \citet{Ferguson2005}
and the high-temperature, Compton-scattering dominated regime by
\citet{Buchler1976}.  Electron conduction opacities are from
\citet{Cassisi2007}. Nuclear screening corrections 
come from \citet{salpeter:54,dewitt:73,alastuey:78,itoh:79}.

\begin{figure}[!htb]
  \centering
  \includegraphics[width=0.5\textwidth]{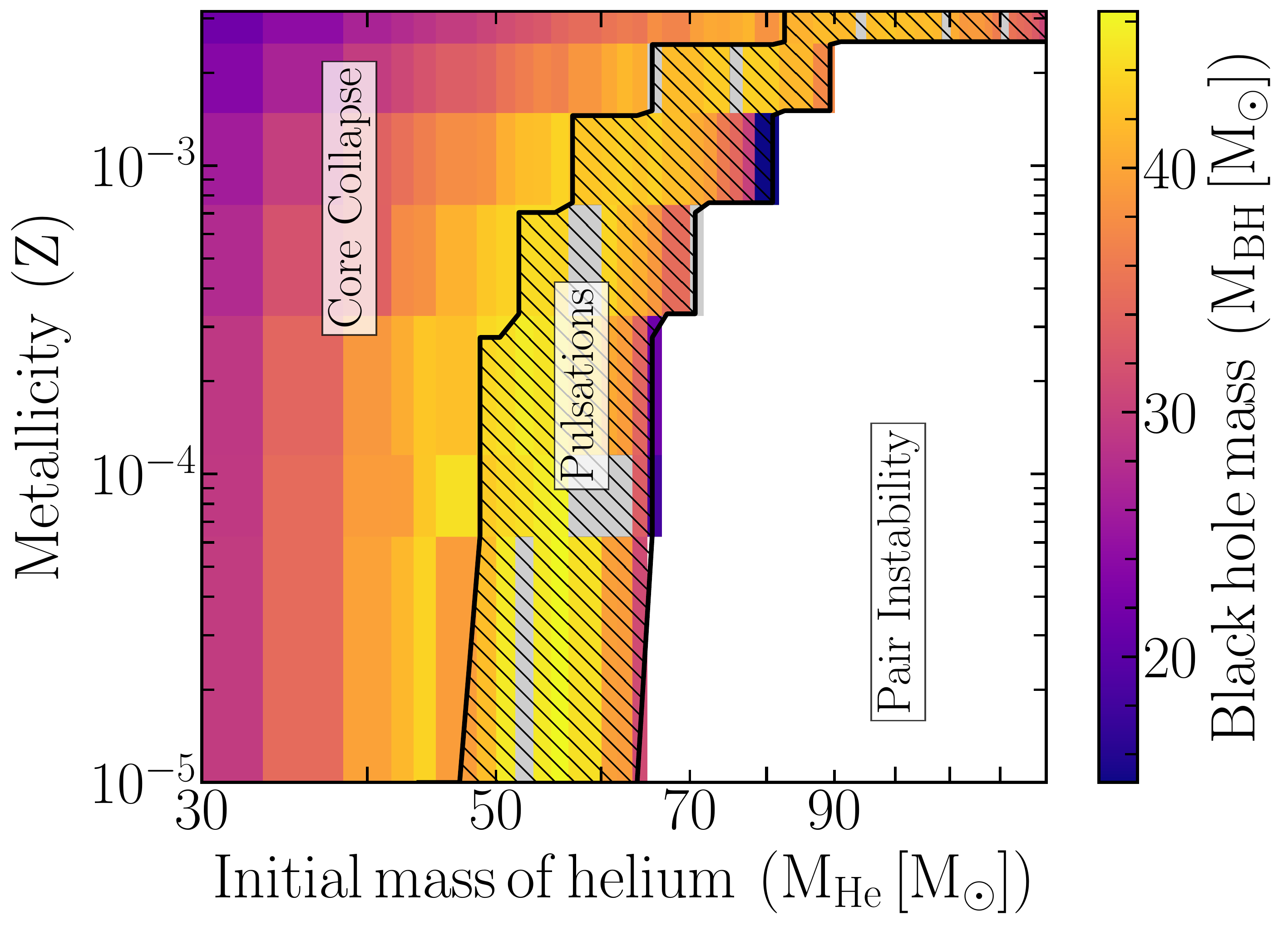}
  \caption{The mass of the BH formed as function of the metallicity of the 
  star and its initial helium star mass. The hashed region
  denotes those models which undergo pulsational mass loss.
   Grey regions indicate models which do not reach CC due to numerical issues. 
   }
  \label{fig:bh_z}
\end{figure}

\section{Robustness of the gap to metallicity}\label{sec:met}

Figure \ref{fig:bh_z} shows the predicted mass of the BH formed from a helium star
with a mass between 
between $30-100\msun$ and initial metallicities between $Z=10^{-5}$ and 
$3\times10^{-3}$. 
At first, as the helium core mass increases, so does the resulting BH mass
 due to the larger initial mass of the star. However, once the star enters the 
pulsational regime, it begins to lose
mass and eventually the amount of mass loss via pulses is sufficient to 
lower the final BH mass. This turn over occurs due to changes in the behavior of the PPI 
pulses.
As the core mass increases, the pulses decrease in number but become more energetic, 
driving off
more mass in each pulse. At the edge of the PISN region, 
the helium cores 
can lose
$\approx10\,\msun$ of material in a single pulse. 

As the core mass is increased further, the first pulse becomes
energetic enough for the star to be completely disrupted in a PISN.
At the lower edge of the BH gap, the most massive helium stars under going 
PPI mass loss without being disrupted lose several tens of solar masses
of material per pulse, leaving behind BHs of $\approx15\msun$.
The lowest mass a BH may have, after undergoing PPISN, is set by the  production 
of \nickel{} inside the star. As the initial mass of the star increases, more 
\nickel{} is produced inside the star. Eventually sufficient \nickel{} is produced
to unbind any material that was not initially driven
away the pulses \citep{marchant:18}. However the exact edge of the 
PPISN/PISN 
boundary, and thus
the minimum BH mass produced by PPISN, is not resolved given our grid spacing.

\begin{figure*}[htp]
  \centering

  \includegraphics[width=1.0\textwidth]{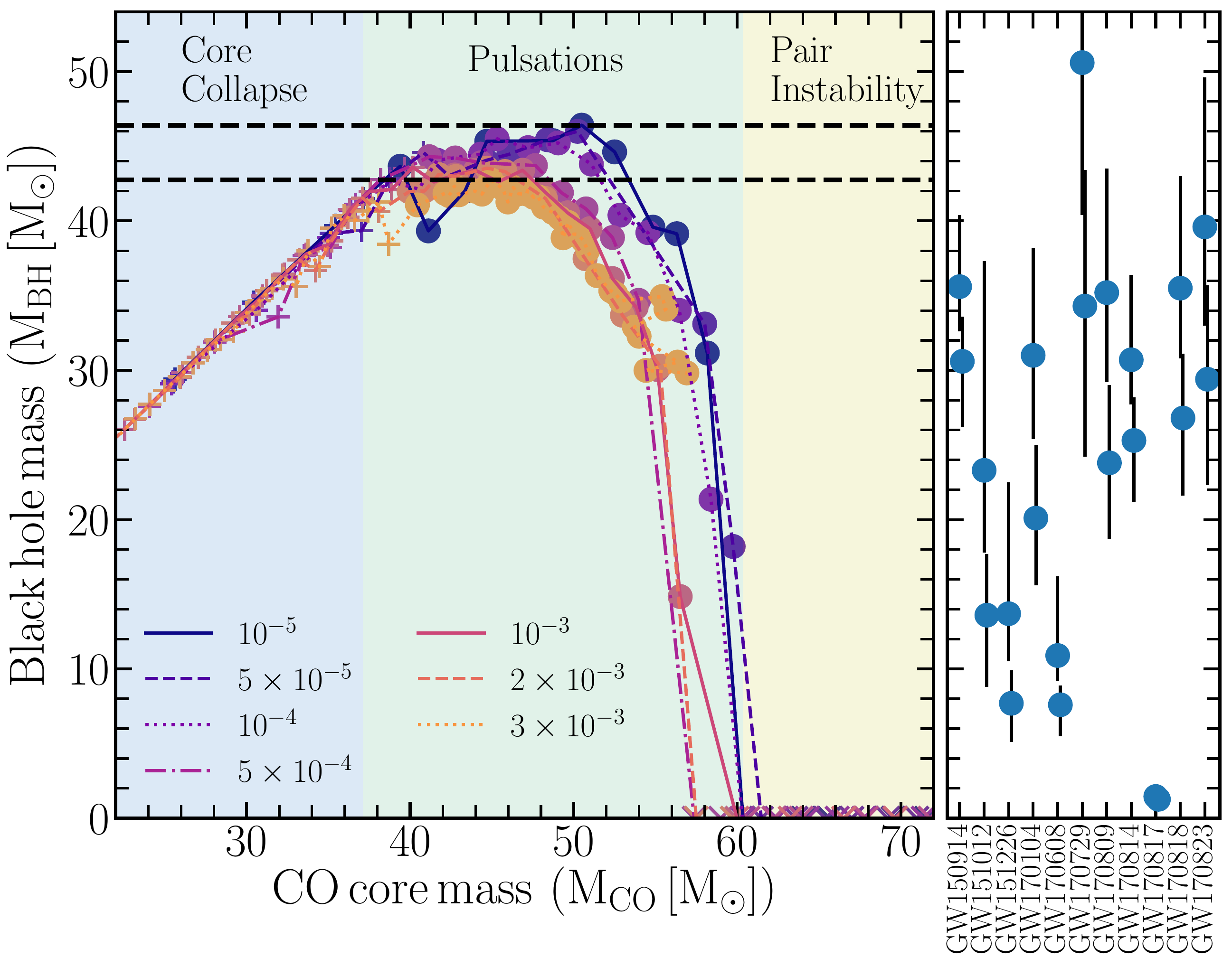}
  
  \caption{Mass of final BH as a function of the CO core mass,
    for different metallicities. Circles denote models that underwent at 
    least one pulse, pluses 
    evolved to directly CC, and crosses undergo a PISN. The left blue region 
    denotes where models undergo CC, the middle green region denotes PPISN,
    while the right yellow region denotes PISN, as determined by stars with 
    $Z=10^{-5}$. Points in the right panel show the current median mass estimates
    for the double compact objects detected by LIGO/VIRGO with their 
    90\% confidence intervals \citep{ligo18b}. Dashed horizontal lines emphasize the maximum spread in the locations for the
    edge of the BH mass gap, or in other words the spread in the maximum BH mass below the PISN BH mass gap.}
    \label{fig:bh_comass_z}
\end{figure*}

As the initial metallicity of the star increases the mass of the BH 
decreases, for fixed initial helium core mass. This is due to increases
in the amount of mass lost via winds before the star enters the PPI
region, which decreases the final mass of the star before collapse.
The progressive shift to the right of the hatched region in 
Figure \ref{fig:bh_z} with increasing Z shows that the
 minimum (and maximum) initial helium core mass needed to undergo pulsations also 
increases as the metallicity increases.
Models with $Z\le3\times10^{-3}$ fully populate the regime for pair instability pulsations.
Numerical limitations prevent us from fully populating the PPISN/PISN region at higher metallicities.
Thus it may be possible to have PPISN at higher metallicities, however this would require helium cores initially more massive
than $\approx100\,\msun$.
Again this is due to the winds; as the winds become stronger we require an
initially more massive progenitor star to retain 
sufficient mass to undergo pulsations.

Figure \ref{fig:bh_comass_z} shows the BH mass as a function of the CO core 
mass over our metallicity range. Here we see a much
tighter relationship between the CO core mass and the final BH mass than in 
Figure \ref{fig:bh_z} between the initial helium core mass and final BH mass.
We find strong PPI pulses removing significant amount of mass between 
CO core masses $\mco\approx38\,\msun$ and
$\mco\approx60\,\msun$.
The upper edge of the PPISN region slightly decreases to $\mco=56\,\msun$ as the
metallicity increases. 
The most massive BHs come from stars with $\mco\approx50\,\msun$, 
not from those with the most massive CO cores that undergo a PPI (in 
Figure \ref{fig:bh_z} these are  $\mco\approx60\,\msun$). This
is due the pulses becoming stronger and thus driving more
mass loss.

We attribute the differences arising from changes in metallicity primarily due to the differences 
in wind mass loss rate. Higher metallicity stars have higher 
wind mass loss rates, which increases the amount of mass loss \citep{castor75,vink:01}. This increased
mass loss forces the convective core to recede, leaving behind a smoother
composition gradient in the outer layers of the star.  
At the highest metallicities the stellar winds have also removed all 
remaining helium from the star
and have begin ejecting C/O rich material, pre-pulses. Thus these 
progenitors would likely look like carbon or oxygen rich
Wolf-Rayet (WC/WO) stars before pulsating. This justifies our choice
of using the CO core mass over the He core mass as a better proxy for the final BH masses.
We note that while the CO-BH mass distribution
is relativity constant over the metallicities considered here, the BH 
formation rate, and hence the merger rate,
will vary as a function of metallicity. This is due to
changes in the initial stellar mass needed to form such massive CO cores. 

The right panel of Figure \ref{fig:bh_comass_z} also shows a comparison 
with the LIGO/VIRGO BH masses detected by the end of the second observing run \citep{ligo18a,ligo18b}. We 
find that the most massive BH LIGO/VIRGO has so far detected is 
consistent with the upper edge of the BH masses we find. 
This is due in part, to the large 90\% confidence intervals on the 
individual BH masses from GW detections. Nevertheless, 
even when considering the much better determined chirp mass of
 GW170729, it remains within the maximum chirp mass predicted 
 assuming random pairing of BHs with mass ratio $\rm{q=M_2/M_1>0.5}$ 
\citep{marchant:18}.

\begin{figure}[htp]
  \centering
  \includegraphics[width=0.5\textwidth]{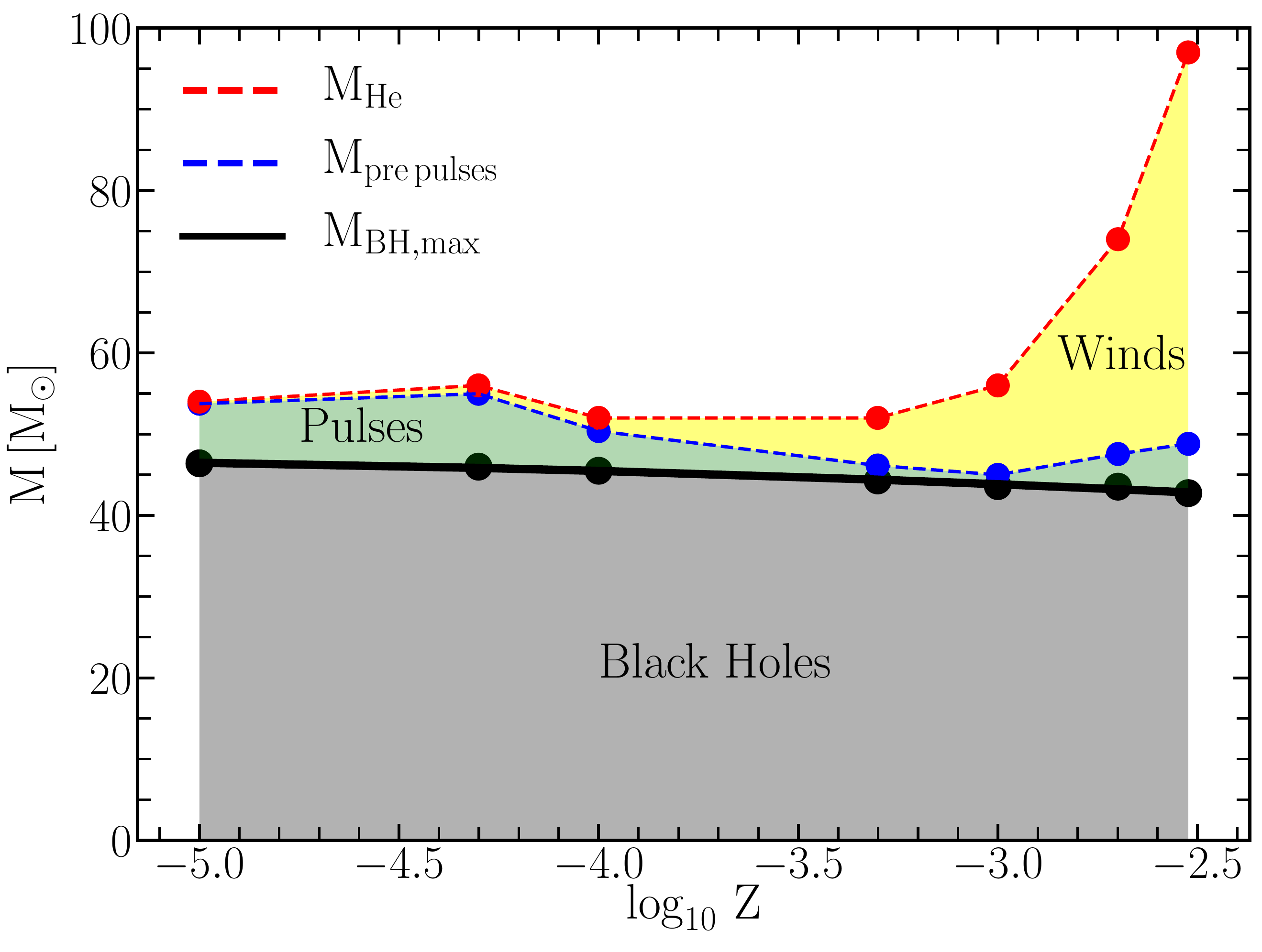}
  \caption{Fate of the mass of the progenitors of the most massive BH below the PISN BH mass gap as a function of metallicity.
    The gray region denotes mass which becomes the BH.
    The green region denotes mass that is lost via pulsations. The yellow region denotes
  mass loss via winds. Black points denote the final BH mass. Blue points denote the edge between
  mass loss in pulsations and in winds. Red points mark the initial helium
  core mass of the star.
  }
  \label{fig:max_bh_z}
\end{figure}

Figure \ref{fig:max_bh_z} shows, as a function of Z, what is the final fate of the mass inside the 
progenitor star forming the most massive BH.
At low metallicities, the weakness of the stellar winds results in most of the initial stellar mass
of the star forming the BH. At higher
metallicites wind mass loss is able to drive approximately half of the initial mass away before
the star collapses to form a BH. The stars making the most massive BHs only 
lose $1-5\,\msun$ of material in the pulsations.

Our models span over 2.5 orders of magnitudes in metallicity, but over such a wide range the maximum
BH mass decreases only slightly between $\rm{M_{BH,Max}}=43-46\,\msun$. This corresponds to 
a $7\%$ variation over the metallicity range considered here, for BHs whose progenitor underwent a PPISN.
The initial helium core mass which forms the 
most massive BHs at each Z
 increases from $\approx54\msun$ at 
$\rm{Z}=10^{-5}$ to $100\,\msun$ at $\rm{Z}=3\times10^{-3}$.
This increase in mass is not due to changes 
in pulse behavior, but instead to the 
increased mass loss due to winds (seen as the yellow shaded region in
figure \ref{fig:max_bh_z}). Thus with a change of only 
6\msun{} in BH mass, the initial mass needed to produce the BH changes
by $\approx50\,\msun$ due to changing the metallicity over 2.5 orders
of magnitude.

\section{Physics dependence of the gap}\label{sec:physics}

\begin{figure*}[ht]
  \centering
  \subfigure[Winds]{\label{fig:bh_wind_pres}\includegraphics[width=0.49\linewidth]{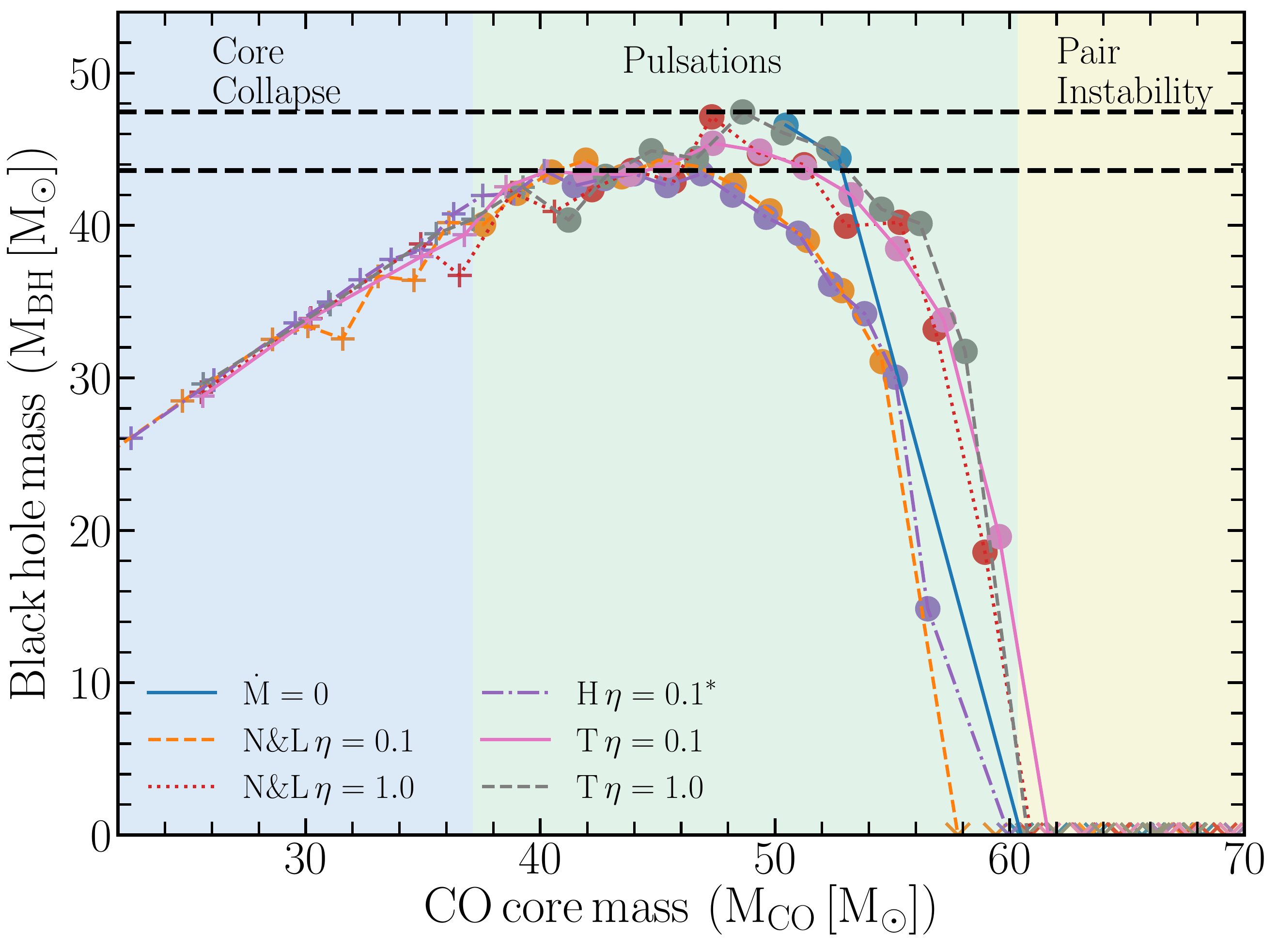}}
  \subfigure[Neutrinos]{\label{fig:bh_neu}\includegraphics[width=0.49\linewidth]{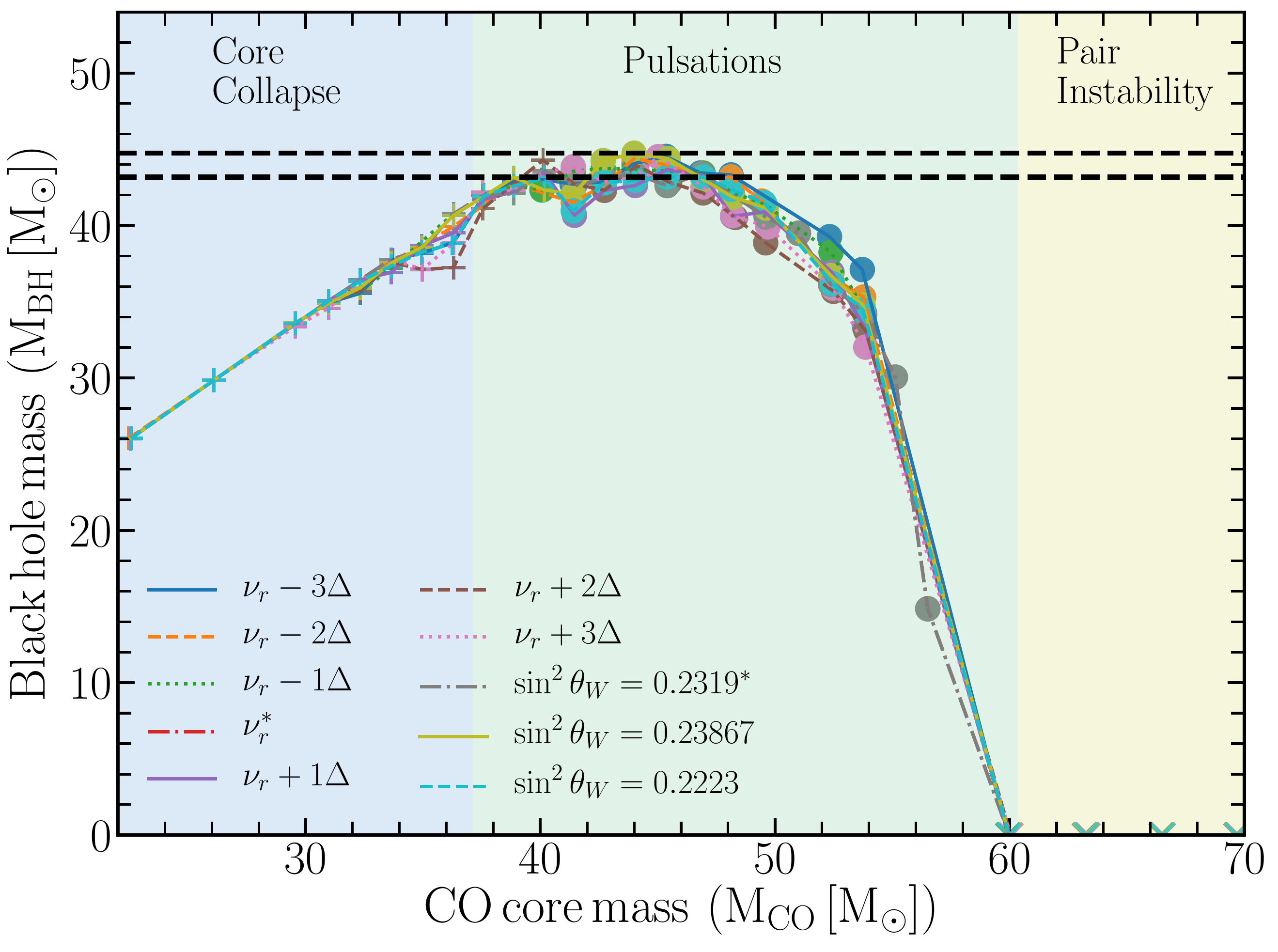}}

  \subfigure[Mixing]{\label{fig:bh_mixing}\includegraphics[width=0.49\linewidth]{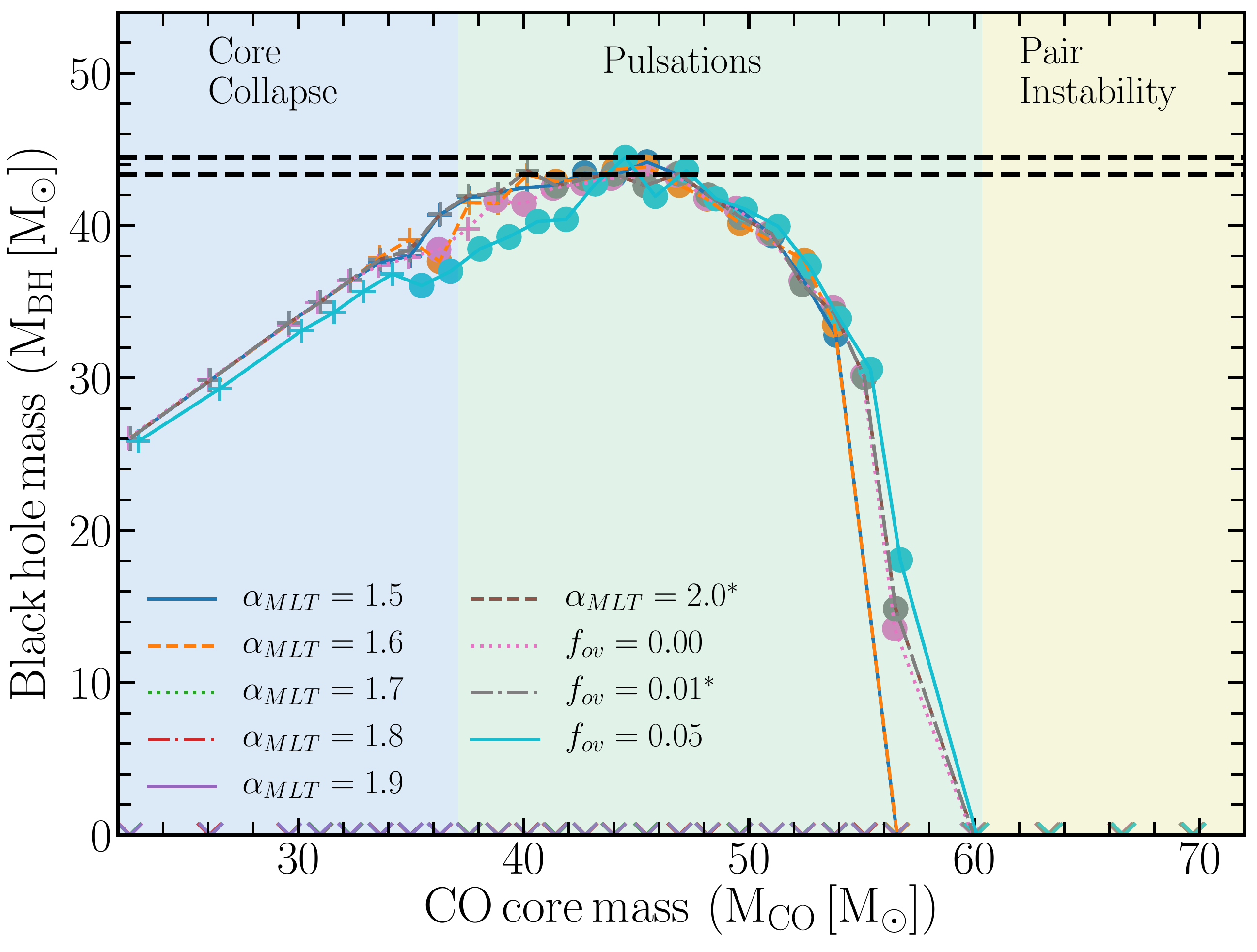}}
  \subfigure[Nuclear reaction rates]{\label{fig:bh_rates}\includegraphics[width=0.49\linewidth]{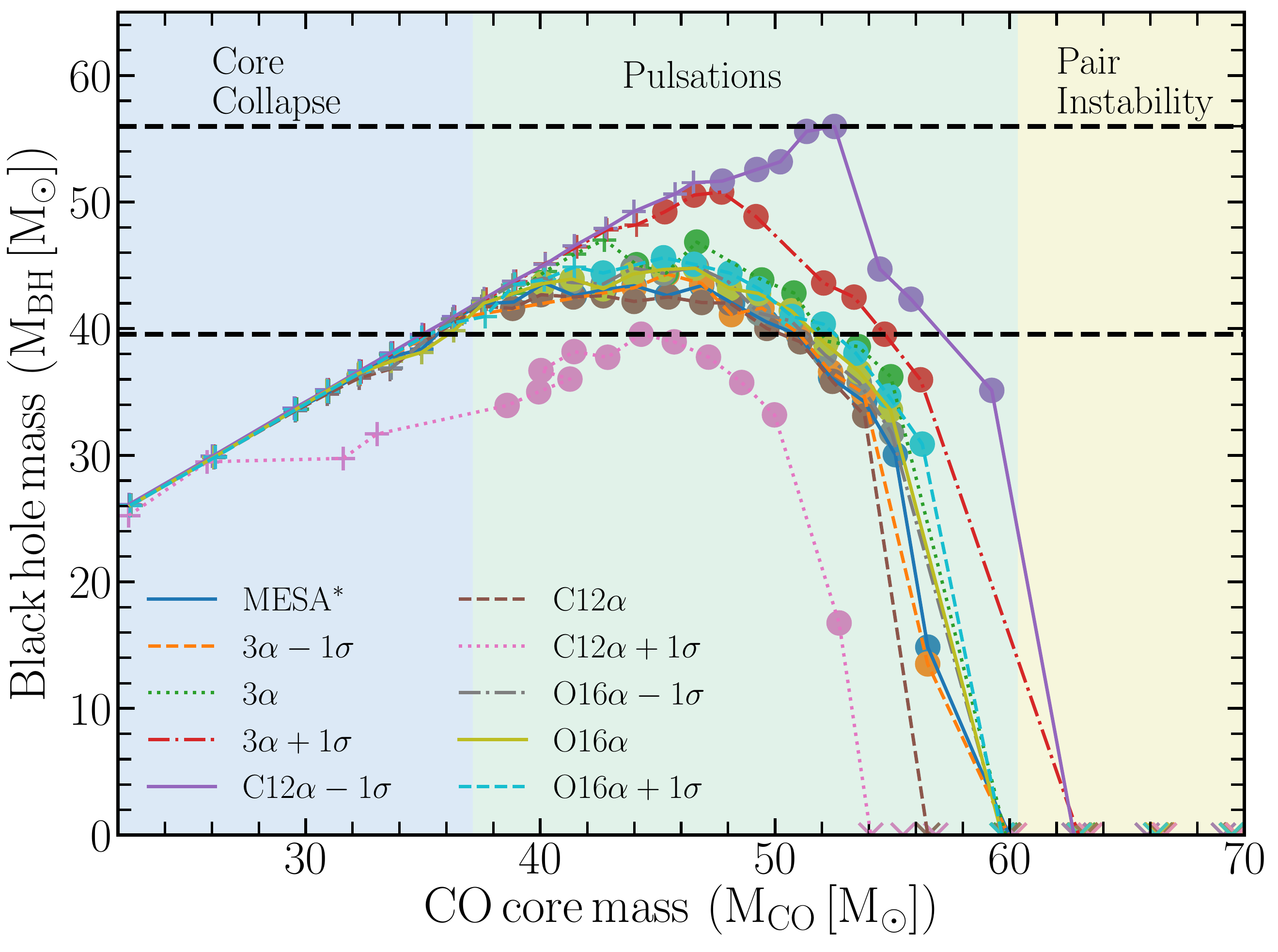}}
  
  \caption{BH mass as function of CO core mass for different physics assumptions. 
  Panel a shows variations in the wind mass loss prescription;
  H is the prescription of \citet{hamann:98}, N\&L is 
  the prescription of \citet{nugis:00}, and T is from \citet{tramper:16}, while
  $\eta$ varies between
  0.1 and 1.0. Panel b shows variations in the neutrino physics; due to the numerical
   uncertainties in the fits \citep{itoh96}, each $\Delta$ represents a scaling of this fitting error;
   and the Weinberg angle. Panel c shows variations in the convective treatment, with 
  varying MLT scale heights $\alpha_{MLT}$ and 
  convective overshoot values $\fover$. Panel d shows variations in a select 
  set of nuclear reaction reactions; \MESA's default rates are from \NACRE{} \citep{angulo99} and \REACLIB{} \citep{cyburt10},
  while the other rates come from \STARLIB{} \citep{sallaska13} as either the median or 
  $\pm1\sigma$ uncertainties, $3\alpha$ is the triple alpha reaction, 
  $\rm{C12\alpha}$ is the $\rm{^{12}C\left(\alpha,\gamma\right)^{16}O}$ 
  reaction, and $\rm{O12\alpha}$ is the 
  $\rm{^{16}O\left(\alpha,\gamma\right)^{20}Ne}$ reaction. Plot 
  symbols have the same meaning as in Figure \ref{fig:bh_comass_z}.
  A star represents our default model assumptions for each physics variation. Dashed lines indicate the range of locations for the
    edge of the BH mass gap.
    Colour 
  shading shows the regions
  between the CC, PPISN, and PISN outcomes for our fiducial set of physics assumptions.}
  \label{fig:bh_param}
\end{figure*}

In figure \ref{fig:bh_param}, we show the variations in the BH mass distribution 
for multiple assumptions of stellar physics, varied within either their theoretical
or experimentally derived uncertainties. Each model is computed at a 
fixed metallicity of $\rm{Z}=10^{-3}$,
with only one parameter varied in each model.

\subsection{Wind prescription}

Figure \ref{fig:bh_wind_pres} shows the effect of different mass loss 
prescriptions 
on the 
CO-BH mass distribution.
Overall the difference in masses between the different prescriptions 
(and $\eta$ values)
is small. The different prescriptions bifurcate into two groups, 
those where $\rm{M}_{\rm{BH,max}}\approx44\msun$ (H$\eta=0.1$ 
and N\&L$\eta=0.1$) and those with $\rm{M}_{\rm{BH,max}}\approx48\msun$ 
($\dot{\rm{M}}=0.0,\rm{N\&L}\eta=1.0$, and $\rm{T}$ (with both $\eta 's$)). 
The models producing
smaller maximum BH masses, also shift their transition to PISN to smaller
 CO core masses. These models
lose more mass via winds and come from $\mheint\approx64\msun$.
The second group, which make $\rm{M}_{\rm{BH,max}}\approx48\msun$ ,
come from $\mheint\approx58\msun$ cores and lose less mass via winds. 
As the
strength of mass loss increases, either though changing the 
wind prescription or increasing the metallicity, the CO-BH mass distribution 
flattens and decreases the 
maximum BH mass. There is no set of models (H) with $\eta=1.0$ shown, as
the amount of mass loss when using this prescription is sufficient 
that no model enters the pulsation region.

\subsection{Neutrino Physics}

Figure \ref{fig:bh_neu} shows the BH mass as a function of the CO core mass 
for variations in the neutrino
rate and the Weinberg angle. Over the rage of variations in neutrino rates considered here,
the effect on the maximum BH mass is small.
As the rate increases we find 
little change in the BH mass distribution, with the maximum BH mass varying by
$\approx1\msun$ and a trend for less massive BHs as the neutrino rate increases.
As the Weinberg angle varies, again the CO-BH mass function is approximately 
constant. Smaller Weinberg angles result in a slightly lower maximum BH mass, 
with a variation of $\sim1.5\msun$ for the range of $\wein$ considered here.

\subsection{Convective mixing}

Figure \ref{fig:bh_mixing} shows variations in \mlt{} between 1.5 and 2.0, 
with our default assumption being $\mlt=2.0$. 
Within these limits there is very little change in the behavior of the BH
masses, with the BH masses slightly decreasing as \mlt{} increases.

Figure \ref{fig:bh_mixing} also shows the effect
 of varying
$\fover$ to be small. The maximum BH mass varies within $1\msun$ over the 
range considered here.
The most significant difference occurs at the PPISN/CC boundary where 
$\fover=0.05$, decreases
the final BH mass relative to the lower \fover{} models.  This is due to a 
change in behavior in
the burning and convection regions at the center of the star. When 
\fover{} is small
the star has a separate off-center and a central burning region, both of which 
drive convection zones.
When \fover{} increases these convection zones can merge, which increases 
the available fuel supply and causes
the pulses to become stronger, driving increased mass loss.

\subsection{Nuclear reaction rates}\label{sec:rates}

Figure \ref{fig:bh_rates} shows the CO-BH mass function for different rates 
computed from \STARLIB{} and our 
default rates from \NACRE{} and \REACLIB. Overall the effect of the 
$\rm{^{16}O\left(\alpha,\gamma\right)^{20}Ne}$ is minimal on 
both the BH mass distribution and the maximum BH mass. However both the $3\alpha$ rate
and the \crate{} rates have a large impact on both the BH mass distribution 
and the maximum BH mass formed.

As the \crate{} rate decreases the maximum BH mass increases, for $+1\sigma$ we find 
$\maxbh=40\msun$ while at
$-1\sigma$ we find $\maxbh=58\msun$. Thus within the 68\% confidence interval for the 
$\rm{C12\alpha}$ the maximum  BH mass varies by $\approx18\msun$ \footnote{For \crate{} reactions with the $+1\sigma$ rate, 
we burn sufficient $^{12}\rm{C}$ during core helium burning such that
we never trigger the CO core mass definition in section 
\ref{sec:method}. Thus we relax our CO core mass definition to be the mass coordinate at 
the maximum extent of the 
core helium burning convection zone.} . 
The median \crate{} rate from \STARLIB{}, from \citet{kunz02}, is smaller than 
the \NACRE{} rate,
thus \STARLIB{} predicts a more massive maximum BH mass. 
\citet{deboer17} also provide an updated \crate{} rate
which is smaller, over the core helium burning temperature range, than \NACRE{}.
Models with this rate showed a similar increase in the maximum BH mass.

As the $3\alpha$ rate, from \citet{angulo99}, increases the
maximum BH mass also increases. This correlates with the 
\crate{} rate behavior; as $3\alpha$ rate increases or the \crate{} rate decreases we increase
the mass fraction of $^{12}\rm{C}$ in the core. For the values tested 
here, this increases from $\approx10\%$ to $\approx30\%$.

We find that as the mass fraction of carbon increases in the core
 the maximum 
BH mass also increases, and also alters the behavior of the pulses.
Higher carbon fractions decrease the range in CO core 
mass within which models undergo pulsations. This would translate into a smaller predicted rate
of PPISN in the Universe, as there is a smaller range of possible progenitors. 
Increasing the mass fraction of carbon also
decreases the fraction of models with strong pulsational mass loss, by weakening the pulses
such that they do not eject mass.
As the carbon fraction increases the BH 
mass distribution sharpens (similar to what is seen with no mass loss in
Figure \ref{fig:bh_wind_pres}). This also shifts the boundary 
between CC/PPISN and between PPISN/PISN to higher masses as the
carbon fraction increases. Moving the boundary between PPISN/PISN to higher CO core masses
would translate to needing a more massives initial star, and thus this would decrease the predicted
rate of PPISN and PISN.

We performed additional tests varying the $\rm{^{12}C+^{12}C}$ 
and $\rm{^{16}O+^{16}O}$ reaction rates\footnote{In the \code{approx21.net} nuclear network
these reactions rates are compound rates where the different output channels
have been combined.} between
0.1 and 10 times their default \MESA{} values as \STARLIB{} does 
not have temperature dependent uncertainties for them. These rates
showed variations in the maximum BH mass of $\sim4\msun$, with the $\rm{^{12}C+^{12}C}$ 
having a larger effect on the maximum BH mass.

Due to the sensitivity of the maximum BH mass to the \crate{} rate,  
the measured value of the maximum BH mass (below the PISN mass gap) can be used to place constraints
on the \crate{} rate (Farmer et al, in prep).

\subsection{Model resolution}

\MESA{} has a number of ways to control the spatial and temporal resolution
of a model. Here 
we vary 
\MESA's \code{mesh\_delta\_coeff}, which controls the maximum 
allowed change in stellar properties between adjacent mesh points 
during the hydrostatic evolution,
between 0.8 and 0.3. Decreasing the value increases the 
resolution. This range corresponds to roughly a factor of two increase in the 
number of grid points.
We also vary 
\MESA's adaptive mesh refinement parameters (AMR), which set 
the resolution during hydrodynamical evolution. We vary
\code{split\_merge\_amr\_nz\_baseline} between 6000 and 10000 
and \code{split\_merge\_amr\_nz\_MaxLong} between 1.25 and 1.15, where 
the second values 
denotes a higher resolution. This leads to and increase by a factor of two
in the number of spatial zones during the evolution of a pulse. 

We have also 
varied \MESA's \code{varcontrol\_target}, which sets
the allowed changed in stellar properties between timesteps, between $5\times10^{-4}$ 
and $5\times10^{-5}$, and varied the \code{max\_timestep\_factor} between 
$1.025$ and $1.05$, which sets
the maximum timestep factor by which \MESA{} can increase a timestep.
This leads to an increase of $\approx30\%$ in number of timesteps 
taken. Over the ranges considered here 
we find changes of $\approx1\msun$ in the maximum BH mass.

Over the range of nuclear networks considered here; 
\code{approx21.net}, \code{mesa\_75.net}, \code{mesa\_128.net};
 there is little change
in the BH mass for a given CO mass, by at most $\approx1\msun$. There is a
trend for larger nuclear networks to produce slightly more massive BHs.

\citet{woosley:17} suggest
that PPI systems need large nuclear networks which can adequately 
follow weak interactions, which \code{approx21.net} does
not follow. However both the CO-BH mass relationship,
and the maximum BH mass vary within $\approx1\msun$ over the networks considered here. 
Changing the isotopes evolved will have an effect
on the composition and final structure of the star as well as the 
composition of the ejecta from the pulses.
However, we find that much of the behavior that determines the final BH mass is set
by the conditions at the initial pulse. This is set by the CO core mass 
and carbon mass fraction, both of which are set by core helium burning,
which is not affected by the lack of weak reactions in \code{approx21.net}.

\section{The maximum black hole mass and its implications}\label{sec:implicat}

\begin{figure}[htp]
  \centering
  \includegraphics[width=0.5\textwidth]{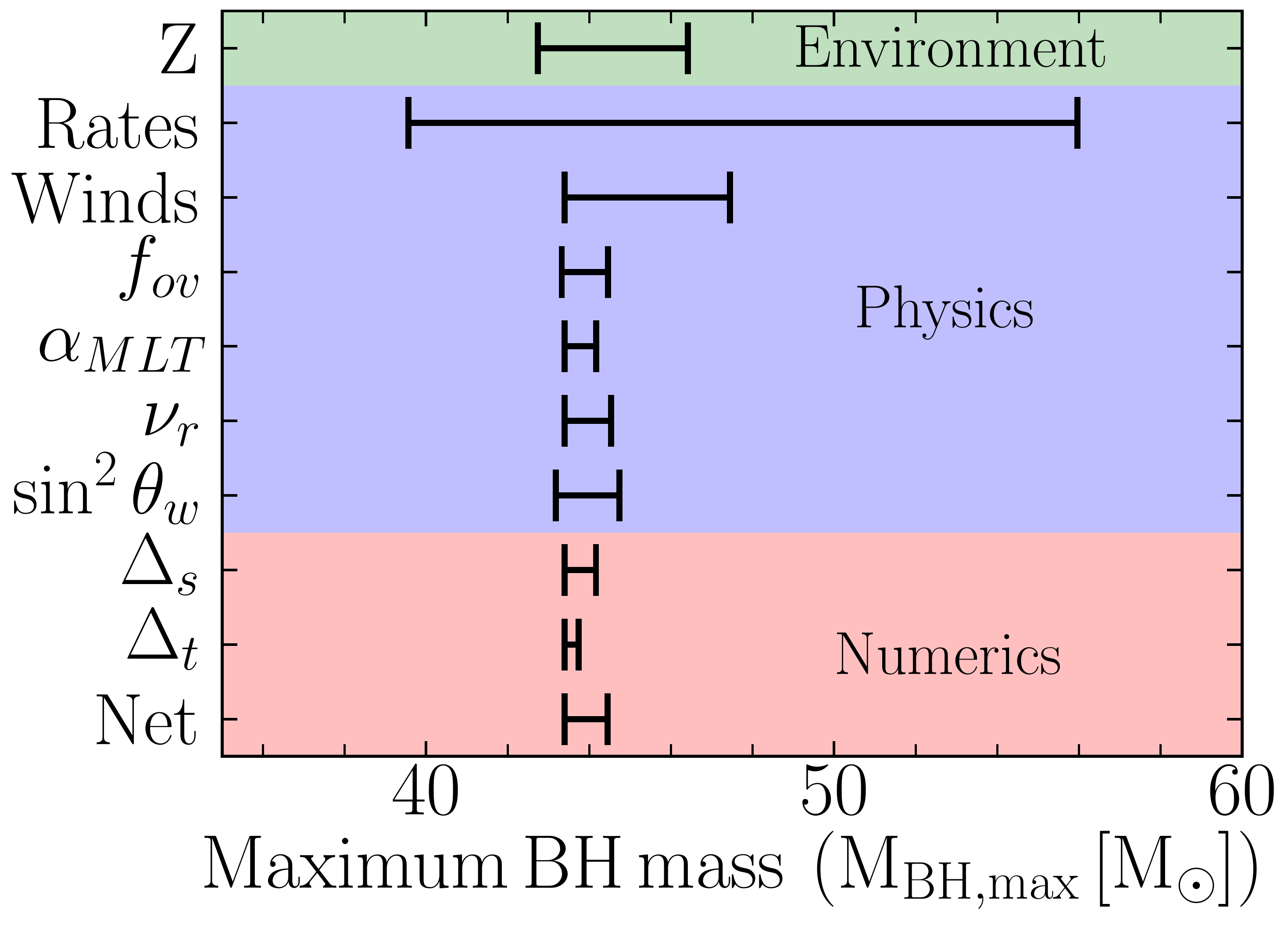}
  \caption{Range of maximum BH masses, for different environment and 
  stellar physics assumptions. See Figure \ref{fig:bh_comass_z} for the range of 
  metallicities considered here, see Figure \ref{fig:bh_param} for the 
  ranges of each physics assumption.}
  \label{fig:bh_massrange}
\end{figure}

Figure \ref{fig:bh_massrange} summarizes the range in the maximum BH mass below the PISN gap
due to the variations
considered in sections \ref{sec:met} and \ref{sec:physics}. These include
those affected by the environment (metallicity) and thus vary across the Universe, 
those for which we have incomplete or uncertain physics 
(rates, winds, $\mlt$, $\fover$, $\neu$, and $\wein$) but we expect to be
 constant in the Universe,
and those that are model dependent (spatial, temporal, and nuclear network resolution). 
For most of the physics for which we are uncertain ($\mlt$, $\fover$, $\neu$, and $\wein$) and 
the model resolution (spatial, temporal, and in number of isotopes) 
there is a limited effected on the maximum BH mass.
These terms place $\approx2\msun$ uncertainties on the maximum BH mass, over the ranges considered here,
contingent on how the different uncertainties are combined.

The next most 
significant factors are the metallicity and winds. We consider these together, since the 
metallicity dependence of wind mass loss rates 
introduces a degeneracy between these two elements.  As we
observe a population of BHs from different progenitor stars with 
varying metallicities, 
then this 7\% variation in 
the maximum BH mass places a minimum level of uncertainty 
on what we can learn from 
the most massive BHs
detected. Given a sufficiently large population of binary BHs (at multiple redshifts)
it may be possible to disentangle the effects of the star formation and
metallicity evolution of the Universe on the BH population \citep{dominik:13,dvorkin16}. However this 
uncertainty which varies over the Universe, is small compared to the current measurement
uncertainties. 

From a gravitational-wave detection we can infer the luminosity distance to the source.  
We also obtain the chirp mass in the detector frame, i.e., the redshifted true chirp mass.   
Knowledge of the true source mass would therefore also provide the redshift to the source, 
and so allows use of gravitational-wave events to measure the expansion history of the 
Universe without the need for electromagnetic detections to supply the redshift of the event.  
Knowledge of the edge of the PISN BH mass gap allows BH mergers to act as ``standardizable sirens'' 
for cosmology (demonstrated by \citet{farr19}, following \citet{schutz:86,holz15}). The sharper the
edge of the PISN mass gap is, the smaller the uncertainty in the derived cosmological parameters that can be achieved \citep{farr19}

The most significant physics variation considered here is due to the nuclear 
physics uncertainties,
and primarily due to the \crate{} rate, leading to a 40\% variation in the 
maximum BH mass. Models having lower
\crate{} rates lose less mass in pulsations and thus
produce more massive BHs. Thus, even with a lack of knowledge about
the environment in which any individual BH
formed, we can still use the detection of sufficiently massive BHs to 
constrain nuclear physics. The most massive detected BH indicates the maximum value
for the \crate{} rate over the core helium burning temperature range.

Given the sensitivity of the BH mass to the CO core mass, the maximum BH mass
formed is effectively independent of its stellar origin. Assuming that both chemically homogeneous
evolution or common envelope evolution can produce a sufficiently massive, H-poor, He core
we would expect those evolutionary scenarios to merging BHs to result in similar final BH masses. 

\section{Comparisons to other work}\label{sec:desc}

In \citet{yoshida:16} they studied PPISN from stars with initial masses between 
140 and 250\msun{} and Z=0.004.
They find the final masses of their stars to be between $50$ and $53\msun$ at 
collapse, broadly consistent with
the masses we find. For our models at Z=0.004 we would expect slightly 
smaller BHs, due to the winds
stripping the outer CO layers of the stars. Another possible source of 
differences may be the choice of the \citet{caughlan88} \crate{} rate \citep{yoshida11}.

Our models agree with the wind-less, metal-free, helium-core models of \citet{woosley:17},
who finds a maximum final BH mass $48\msun$. This agrees with our wind-less, 
models where we also find a maximum BH mass of $48\msun$ (though we evolve them at a non-zero metallicity). 
\citet{woosley:17} also find
a maximum BH mass of $52\msun$ for models which did not remove their entire hydrogen 
envelope. Although they are not directly comparable to our results, which assume all helium has been removed,
they provide bounds on the variation in the maximum BH mass, if the H-envelope is not completely removed, of
$\approx4\msun$.
\citet{woosley19} investigated the evolution of naked He cores finding a maximum 
BH mass below the gap of $46\msun$ in agreement with our results, for our default \crate{} rate and
assumptions on metallicity.

\citet{takahashi18} looked at the effect of varying the \crate{} rate on the 
boundary between PPISN and PISN. Our results
are not directly comparable as they used the \citet{caughlan88} rate, 
while by default we use the \NACRE{} rate. 
They found that as the \crate{} rate decreases,
which increases the $^{12}\rm{C}$ fraction, the PISN boundary shifts to higher 
initial masses, similar to our findings. \citet{takahashi18} also
find a narrowing of the PPISN region (in initial mass space), consistent with our results.

\citet{leung19} studied the evolution of helium cores between $80-140\msun$ and 
$\rm{Z}=10^{-3}-1.0\rm{Z_{\odot}}$,
with a previous \MESA{} release (version 8118) and a different treatment of the hydrodynamics. Their
results are in agreement with ours. They find a maximum BH mass of 
$\approx50\msun$, which is larger than what we predict, likely due to their lowest metallicity
models ($\rm{Z=10^{-2}Z_{\odot}}$) having no wind mass loss. Our no 
mass-loss models at $\rm{Z=10^{-3}}$ have a maximum BH mass of $48\msun$. At higher
metallicities we find that wind mass loss is sufficient to make the maximum BH come from
a CC event rather than a PPISN. The remaining
differences may be due to other differences in choice of parameters and/or 
changes between the two \MESA{} versions.

In this work we did not consider the effects of rotation on the stellar progenitors, 
which is especially important
if they came from chemically homogeneous evolution \citep{demink:09,demink16,marchant:16}. However,
\citet{chatzopoulos:12,chatzopoulos:12b} have previously studied the impact of 
rotation on zero metallicity models evolved first with \MESA{} and then \code{FLASH}, in 1-D, 
for the dynamical evolution. They found that increasing the rotation rate the initial mass
for PPISN or PISN decreases. This is due to the increase
in CO core mass due to rotational mixing. They find PPISN from stars with core
masses between $\approx40-65\msun$, which agrees with our results. However the
impact of centrifugal support on the resulting BH masses from PPISN 
is an open problem. Rotation may also affect the final BH mass depending on how the BH
is formed and whether material with high angular momentum is accreted into the BH. 
This may, however, be more relevant for the final spin of the BH compared to the final mass \citep{rockefeller:06,fryer04,batta17}.

\section{Summary and Conclusions}
\label{sec:conc}

The prediction of a gap in the mass distribution of BHs from the 
collapse of stars dates back to the sixties, when the theory of
pair instability evolution had first been developed \citep{fowler:64, barkat:67}. However, it 
is only recently that the possibility of testing this prediction 
directly with gravitational waves has opened. 
As the presently observed population of binary BHs is compatible with 
having stellar origin \citep[][]{ligo18a,ligo18b}, instead of dynamical or primordial, 
we can use stellar evolution models to interpret the upper 
end of the BH mass distribution.

We find that the evolution of single bare He cores robustly predicts a 
maximum BH mass of $\approx45\msun$, and that
this value is relativity insensitive to variations in the input physics, the 
algorithmic approach,
and the metallicity of the models. In particular, despite the uncertain 
wind mass loss rates of 
massive stars, we find a variation of the maximum BH mass of only 
$\approx7\%$ (from $\approx43\msun$ to $\approx46\msun$)
over 2.5 orders of magnitude in metallicity.
This implies that detailed knowledge of the host galaxy of merging 
binary BHs 
is not required to use gravitational wave detections to probe
the physics of the unobserved stellar progenitors.

The insensitivity to metallicity of the maximum BH mass below the gap
 might also allow for cosmological applications.
If its value can accurately be determined, it can provide a ``standard siren'' \citep{schutz:86,holz15,farr19},
allowing estimates of both a redshift and a luminosity distance to the mergers
from just the gravitational wave detection.

Assuming a stellar origin, the most massive BHs detected below 
the pair instability mass gap
might be used to further constrain nuclear physics, specifically the 
\crate{} reaction in the core helium burning regime. 
In particular, the maximum BH mass puts an
upper limit on this reaction rate. Other physics variations including 
neutrino physics, wind algorithms, and
chemical mixing have sub-dominant effects on the maximum BH hole mass and 
negligible contributions to the uncertainty
compared to the typical observational uncertainties.

We note however that our estimates of the BH mass may be over-predicted 
if, for instance, a significant amount of mass is loss via
neutrinos during the final collapse \citep{coughlin:18}. Also, 
our simulations do not account self-consistently for binary 
interactions between the progenitor stars,
which deserves further attention \citep{gotberg:17,marchant:18}.

If BHs with masses inside the predicted PISN mass gap are
detected they
could either have non-stellar origins, be the result of multiple mergers in a cluster \citep{rodriguez16a, stone:17,dicarlo19},
or if the star was H-rich at the time of collapse and then merged in a cluster \citep{vigna19,spera2019}.
However,
the expected rate of mergers that include a BH in the mass gap is small,
due to the requirement that the BHs be the result themselves of previous mergers and
that they stayed bound to the cluster after the merger \citep{gerosa19}. Whether
BHs are ejected from clusters also depends strongly on whether they are born spinning;
if they do not spin then they are more likely to stay bound in the cluster \citep{rodriguez19}.
If the star can retain its H-rich envelope, by evolving as an isolated single star and then
merging with another BH in a dense cluster, then it might be massive enough to enter the mass
gap \citep{woosley:17}. However, the expected rate of merges in dense clusters is a factor 40
less than that of isolated binary mergers \citep{rodriguez16a,belczynski:16nat}. 

The present and upcoming detections of binary BH mergers might provide 
evidence constraining the death of the most massive stars before we 
might be able to
might be able to
unequivocally observe these phenomena in the electromagnetic spectrum 
\citep{stevenson:19}. Our results suggest, that with a large population of
merger events, we
can put constraints on uncertain nuclear physics and provide a 
new tool for cosmology. 

\begin{acknowledgements} 
We acknowledge helpful discussions with W.~ Farr, D.~Brown, B.~Paxton, F.~Timmes, and I.~Mandel, 
Y.~G\"otberg, D.~Hendriks, E.~C.~Laplace, E.~Zapartas.
RF is supported by the Netherlands Organization for Scientific Research (NWO) 
through a top module 2 grant with project number 614.001.501 (PI de Mink). 
SdM and MR acknowledge funding by the European Union's Horizon 2020 research 
and innovation programme from the European Research Council (ERC) 
(Grant agreement No.\ 715063), and by the Netherlands Organization for 
Scientific Research (NWO) as part of the Vidi research program BinWaves 
with project number 639.042.728.
SdM acknowledges the Black Hole Initiative at Harvard University, 
which is funded by grants from the John Templeton Foundation and the Gordon and Betty Moore Foundation to Harvard University.
PM acknowledges support from NSF
grant AST-1517753 and the Senior Fellow of the Canadian 
Institute for Advanced Research (CIFAR) program
in Gravity and Extreme Universe, both granted to
Vassiliki Kalogera at Northwestern University.
This research was supported in part by the National Science Foundation under Grant No. NSF PHY-1748958.
This work was carried out on the Dutch national e-infrastructure 
with the support of SURF Cooperative. This research has made use of NASA's 
Astrophysics Data System.
\end{acknowledgements}

\software{
\texttt{mesaPlot} \citep{mesaplot}, 
\texttt{mesaSDK} \citep{mesasdk}, 
\texttt{ipython/jupyter} \citep{perez_2007_aa,kluyver_2016_aa}, 
\texttt{matplotlib} \citep{hunter_2007_aa}, 
\texttt{NumPy} \citep{der_walt_2011_aa}, and
\MESA \citep{paxton:11,paxton:13,paxton:15,paxton:18,paxton:19}.
         }

\appendix{}

\section{Analytic fits for population synthesis}

Population synthesis studies of the impact of PPI on the
gravitational wave mergers distributions have relied on numerical
fits to the results of \cite{woosley:17} expressed as a function of the helium
core mass \citep[e.g.,][]{belczynski:16, spera:17, stevenson:19}. However,
at high metallicites we
find the stars are stripped of all helium, leaving a bare CO core.
As the CO core mass at the time of core-collapse is a quantity that is available in
population synthesis calculations (although possibly defined differently
compared to here, \citealt{hurley:00}), we recommend to use 
\mco~as the independent variable to determine
the final BH mass of stars. Though this only applies to stars
that have lost their hydrogen envelopes 
either in binary interactions or due to wind mass loss.

For any given choice of physics and numerics, the second most important
parameter, after \mco, determining the final BH mass is the initial
metallicity of the star $Z$. We provide an approximate fit to the BH masses in figure \ref{fig:bh_comass_z}
in terms of these two parameters:
\begin{equation}\label{eq:mco_mbh}
\rm{M_{BH}} = \begin{cases}

4 + \rm{M_{CO}} & \rm{M_{CO}}  < 38 \\

\parbox[t][][t]{.3\linewidth}{$a_1\rm{M^2_{CO}}+ a_2\rm{M_{CO}} + a_3\log_{10}(\rm{Z}) + a_4$} & \parbox[t][][t]{.40\linewidth}{$38 \leq \rm{M_{CO}} \leq 60$} \\
0.0 & 60 < \rm{M_{CO}}  \\

\end{cases}
\end{equation}
Where $a_1=-0.096$, $a_2=8.564$, $a_3=-2.07$, and 
$a_4=-152.97$, where all masses are in \msun{} and is accurate to $\approx20\%$, though the fit accuracy
decreases as the metallicity decreases.

We note that for
$M_\mathrm{CO}<38\,M_\odot$ weak pulses that do not result in significant mass
ejection are still possible, and might have an effect on the orbital
properties of a binary system \citep[e.g.,][]{marchant:18}. Moreover,
the fit of~\ref{eq:mco_mbh} does not contain information on the mass lost per
each individual pulse, and on the timing of the pulses, which might both influence the evolution
in a binary.

Another important result of this study is the small sensitivity of the maximum
BH mass below the pair instability gap to metallicity, with only a
$\approx7\%$ variation over a range in $Z$ spanning 2.5 orders of magnitude.
Therefore, the
maximum BH mass might be used as a ``standard siren'' for cosmological
applications once sufficiently large samples of BHs are detected. We also
provide an approximate fit to the maximum BH mass below the pair instability gap
as a function of the metallicity which expresses this weak dependence:
\begin{equation}
\rm{M_{BH,max}} = b_1 + b_2\log_{10}(\rm{Z}) + b_3\left[\log_{10}(\rm{Z})\right]^2
\end{equation}\label{eq:bhmass}
where $b_1=35.1$, $b_2=-3.9$, and 
$b_3=-0.32$, where the resulting $M_\mathrm{BH, max}$ is in solar units and is accurate
to $\approx3\%$.
This can be applied also to metallicities $\rm{Z}<10^{-5}$, lower 
than considered here, since we do not expect significant (line driven) wind mass
loss in this regime. However it is unlikely to be valid
for stars with $\rm{Z}>3\times10^{-3}$, due to their 
stronger winds that prevents the formation of sufficiently massive CO cores to
experience PPI-driven mass loss.

\section{Mass loss from progenitors}\label{sec:table}

Table \ref{tab:data} shows the amount of mass loss and final fate for our fiducial set of stellar parameters.
A full version of the table for all models, is available online, for the other parameters considered here. Table \ref{tab:data} shows:
the initial (helium) mass; the helium and carbon/oxygen core masses, measured before the pulsations begin; the final BH mass; the
mass lost in pulses; mass loss in winds; mass lost
at the final supernovae; and the final fate of the star.
The mass lost at supernovae is a combination of the mass loss due to material having a binding 
energy $<10^{48}\,\ergs$ \citep{nadezhin:80,lovegrove:13,fernandez18} and material that is in
the process of being ejected (i.e., it is moving faster than the local escape velocity) but has not been removed from
the model at the time of core collapse.

\begin{deluxetable*}{cc|cccccccc}
\tablewidth{\linewidth}

\tablecaption{Fate of the mass of the progenitors, for our fiducial model. A full table for all models, is available online\label{tab:data}}

\tablehead{\colhead{Parameter} & \colhead{Value} & \dcolhead{\rm{M_{int}}} & 
\dcolhead{\rm{M_{he}}} & \dcolhead{\rm{M_{co}}} & \dcolhead{\rm{M_{bh}}} & 
\dcolhead{\Delta \rm{M_{\rm{pulse}}}} & \dcolhead{\Delta \rm{M_{\rm{wind}}}} & \dcolhead{\Delta \rm{M_{\rm{SN}}}} & \colhead{Fate} }
\startdata
\multirow{29}{*}{Z} & \multirow{29}{*}{$10^{-3}$}  &        30 &        26.12 &        22.55 &        26.05 &         0.00 &         3.88 &         0.08 & CC  \\
 &  &        35 &        29.94 &        26.09 &        29.85 &         0.00 &         5.06 &         0.10 & CC  \\
 &  &        40 &        33.66 &        29.57 &        33.60 &         0.00 &         6.34 &         0.06 & CC  \\
 &  &        42 &        35.12 &        30.98 &        34.97 &         0.00 &         6.88 &         0.15 & CC  \\
 &  &        44 &        36.57 &        32.32 &        36.43 &         0.00 &         7.43 &         0.14 & CC  \\
 &  &        46 &        38.00 &        33.64 &        37.78 &         0.00 &         8.00 &         0.22 & CC  \\
 &  &        48 &        39.42 &        34.95 &        38.38 &         0.00 &         8.58 &         1.04 & CC  \\
 &  &        50 &        40.83 &        36.30 &        40.76 &         0.00 &         9.17 &         0.08 & CC  \\
 &  &        52 &        42.23 &        37.55 &        41.97 &         0.00 &         9.77 &         0.26 & CC  \\
 &  &        54 &        43.62 &        38.86 &        42.10 &         0.00 &        10.38 &         1.52 & CC  \\
 &  &        56 &        44.99 &        40.16 &        43.60 &         0.00 &        11.01 &         1.39 & CC  \\
 &  &        58 &        46.36 &        41.45 &        42.61 &         3.55 &        11.64 &         0.20 & PPISN  \\
 &  &        60 &        47.71 &        42.73 &        43.08 &         4.33 &        12.29 &         0.30 & PPISN  \\
 &  &        62 &        49.06 &        44.00 &        43.39 &         4.66 &        12.94 &         1.01 & PPISN  \\
 &  &        64 &        50.39 &        45.40 &        42.62 &         6.63 &        13.61 &         1.14 & PPISN  \\
 &  &        66 &        51.73 &        46.88 &        43.40 &         7.83 &        14.27 &         0.50 & PPISN  \\
 &  &        68 &        53.04 &        48.19 &        42.00 &        10.27 &        14.96 &         0.78 & PPISN  \\
 &  &        70 &        54.36 &        49.63 &        40.54 &        12.87 &        15.64 &         0.94 & PPISN  \\
 &  &        72 &        55.70 &        51.00 &        39.49 &        15.24 &        16.30 &         0.97 & PPISN  \\
 &  &        74 &        56.96 &        52.37 &        36.14 &        19.22 &        17.04 &         1.59 & PPISN  \\
 &  &        76 &        58.25 &        53.82 &        34.21 &        23.26 &        17.75 &         0.78 & PPISN  \\
 &  &        78 &        59.55 &        55.13 &        30.05 &        26.74 &        18.45 &         2.76 & PPISN  \\
 &  &        80 &        60.83 &        56.51 &        14.85 &        45.88 &        19.17 &         0.11 & PPISN  \\
 &  &        85 &        64.02 &        59.98 &         0.00 &        64.02 &        20.98 &         0.00 & PISN  \\
 &  &        90 &        67.09 &        63.24 &         0.00 &        67.09 &        22.91 &         0.00 & PISN  \\
 &  &        95 &        70.12 &        66.51 &         0.00 &        70.12 &        24.88 &         0.00 & PISN  \\
 &  &       100 &        73.10 &        69.66 &         0.00 &        73.10 &        26.90 &         0.00 & PISN  \\
 &  &       105 &        76.05 &        72.80 &         0.00 &        76.05 &        28.95 &         0.00 & PISN  \\
 &  &       110 &        78.95 &        75.88 &         0.00 &        78.95 &        31.05 &         0.00 & PISN  \\
\enddata

\end{deluxetable*}

\bibliographystyle{aasjournal}
\bibliography{paper}

\end{document}